\documentclass{article}
\usepackage[utf8]{inputenc}
\usepackage[margin=2.5cm]{geometry}
\usepackage{hyperref}

\usepackage{amsmath,amsfonts,amssymb,amsthm,latexsym,mathrsfs,pdfsync,color,graphicx,dsfont}
\usepackage{algorithm}
\usepackage{caption, subcaption, tabularx, booktabs, pdfpages}
\usepackage{algpseudocode}
\usepackage[english]{babel}
\usepackage{multirow}

\numberwithin{equation}{section}

\newcommand{\T}{\mathcal{T}}
\newcommand{\J}{\mathcal{J}}

\newcommand{\R}{\mathbb{R}}
\newcommand{\E}{\mathbb{E}}

\newcommand{\Q}{\mathbb{Q}}

\title{On Deep Learning for computing the Dynamic Initial Margin and Margin Value Adjustment}
\author{Joel P. Villarino$^{1,2,}$\footnote{Corresponding author: joel.perez.villarino@udc.es}, Álvaro Leitao$^{2,3}$}
\date{%
    $^1$Department of Mathematics, University of A Coruña, Spain\\%
    $^2$CITIC Research Centre, University of A Coruña, Spain\\%
    $^3$Universitat Oberta de Catalunya (UOC), Spain\\[3ex]%
    \today
}

\begin{document}

\maketitle

    \bigskip
    \noindent
    
    {\small{\bf ABSTRACT.}
        The present work addresses the challenge of training neural networks for Dynamic Initial Margin (DIM) computation in counterparty credit risk, a task traditionally burdened by the high costs associated with generating training datasets through nested Monte Carlo (MC) simulations. By condensing the initial market state variables into an input vector, determined through an interest rate model and a parsimonious parameterization of the current interest rate term structure, 
we construct a training dataset where labels are noisy but unbiased DIM samples derived from single MC paths. A multi-output neural network structure is employed to handle DIM as a time-dependent function, facilitating training across a mesh of monitoring times. The methodology offers significant advantages: it reduces the dataset generation cost to a single MC execution and parameterizes the neural network by initial market state variables, obviating the need for repeated training. Experimental results demonstrate the approach's convergence properties and robustness across different interest rate models (Vasicek and Hull-White) and portfolio complexities, validating its general applicability and efficiency in more realistic scenarios.

    }

\section{Introduction} \label{sec:intro}
Due to the financial crisis experienced in 2008, which led to the bankruptcy of some major financial institutions, such as Lehman Brothers, the G8 World Council promoted the regulation of stricter actions for financial markets, specially in the over-the-counter (OTC) derivatives market. These are bilateral transactions between (often large) institutions prevailed under certain opacity, giving rise to a counterparty credit risk (CCR), a risk due to the failure to fulfill contractual payment obligations, as a result of outstanding cashflows that are not yet settled. Since then, financial regulators have introduced the obligation of adopting measures to mitigate CCR, which leads to a proliferation of models that take such risks into account in derivatives pricing. They became globally known as \emph{x}-value adjustment (xVA), where the \emph{x} accounts for the kind of risk to be mitigated. The very first ones were credit and debit value adjustments (CVA and DVA), whose foundations can be seen in, e.g., \cite{brigo2005, brigo2011, brigo2014}.

Following the regulatory requirements to mitigate the systemic risk associated with non-cleared OTC derivatives, specially in the context of interest rate derivatives, another key tool is the so-called collateralisation. It is based on the periodic deposit of an amount of money (or some other security if it is allowed) to serve as a protection against counterparty default. More concretely, the Basel Committee on Banking Supervision (BCBS) and the International Organization of Securities Commissions (IOSCO) mandate institutions engaging in these transactions to post bilateral Variation Margin (VM) and Initial Margin (IM) on a daily basis at a netting set level, \cite{basel3}. VM aims to cover the current exposure resulting from changes in the instrument's mark-to-market value by reflecting its current size, but has an associated funding cost, known as funding value adjustment (FVA), which is reflected by the expected exposure of the portfolio. Theoretical derivation is provided and discussed in \cite{brigo2011, brigo2014}. Such a collateral is not sufficient to fully mitigate the credit risk of the trades, which still arises from the time required for the surviving party to close-out its position after default. This close-out time is known as Margin Period of Risk (MPoR), typically of $10$ business days. Then, IM aims to cover the potential future exposure that could arise in such a period, acting as a protection against pronounced changes in the value of the trades. The BCBS requires that the IM collateral for bilateral OTC derivatives should be at least at the level of a $99\%$ value-at-risk (VaR) changes over the MPoR. Given that the exchange of IM must take place over the life of the transactions, it is to be expected that posting this collateral will generate, as in the case of VM, associated funding costs. The total expected cost is known as margin value adjustment (MVA) and was introduced for the first time by the authors in \cite{Green2014}. The calculation of this adjustment implies the necessity to accurately compute not only the spot IM value, but also to forecast its future levels, a problem known as Dynamic Initial Margin (DIM).  

Traditionally, the use of methodologies based on VaR metrics generates disputes between the parties because of their dependency on the model under consideration. The IM calculation was not exempt, so in order to simplify the reconciliation between counterparties, the \textit{International Swaps and Derivatives Association} (ISDA) has promoted a method for computing spot IM, called \textit{Standard Initial Margin Model} (SIMM) \cite{isda_informe}. This model is designed to estimate a $99\%$ VaR level by leveraging first order portfolio sensitivities (delta and vega) to specific risk factors, with the appropriate weighting based on parameter calibration conducted during stressful periods. Hence, the computation of the MVA under ISDA SIMM approach boils down to the challenge of computing the forward portfolio sensitivities to a large set of underlying risk factors. A detailed explanation of the spot IM-SIMM and the extension to the forward computation can be found at \cite{isdameth, Bianchetti2021}, respectively. In both of them, it is presented the approach commonly known as brute-force simulation. Basically, it is a Monte Carlo (MC) framework where we need to compute, for each MC path, the forward portfolio sensitivities with respect to each SIMM risk factor. This leads to a nested MC simulation which has a high computational cost, specially for exotic derivatives that have no closed-form pricing formula.

In practice, where portfolios can be made up of hundreds or thousands of products, the high computational cost of the aforementioned brute-force approach means that may not be feasible in many production systems. So, in recent years, new solutions have emerged to approximate such quantities. On the one hand, a great deal of work has focused on numerical regression techniques based on the modification and improvement of the least-squares MC approach presented in \cite{longstaff2001}. The authors in \cite{lakhany2021} employ such a numerical method to efficiently estimate the sensitivities required in the IM-SIMM methodology, whilst the authors in \cite{McWalter2022} have proposed a statistical method in which they fit DIM conditional moments, estimated using least-squares MC, via Johnson-type distributions. Other statistical approaches can be found in, e.g., \cite{anfuso2017}. On the other hand, much effort has been put into improving the computing speed of forward sensitivities. In this direction, we can highlight works such as \cite{zeron2020}, where the authors present a method based on Chebyshev tensors to compute dynamic sensitivities of financial instruments within a MC simulation; or \cite{capriotti2017, Antonov3, jain2019}, where it has been shown how adjoint algorithmic differentiation (AAD) can be used to calculate sensitivities (even along the MC paths) reliably and orders of magnitude faster than with finite-difference approaches.

In recent years, new methodologies, based on the advances in machine learning and deep learning, have arisen in the quantitative finance field (see \cite{hovarth2021, Savine2020, liu2019, liu2021}, for example). Accounting for CCR and xVA in general, we highlight works like \cite{Villarino2023}, where the authors employ deep learning to solve partial differential equations for pricing options subject to CCR; \cite{albanese2021}, where they compute xVA quantities combining forward/backward stochastic differential equation (SDE) formulations with deep regression techniques; or \cite{Gnoatto2023}, where a computational framework for portfolio-wise risk management founded on recursive application of a neural network-based BSDE solver is presented. In the particular case of DIM and MVA we also find works using deep learning tools. For example, in \cite{aryan2020} a neural network-based method to approximate the Bermudan pricing function and sensitivities is proposed; in \cite{Xun2019} the authors achieve a neural network-based IM simulation of a portfolio consisting of $50$ trades; or \cite{Hoencamp2024}, where a static replication algorithm for interest rate options with early-exercise features is combined with neural networks to get conditional prices and sensitivities, which serve as an input to SIMM-driven MVA quantification.

However, many of the these deep learning alternatives to compute DIM-SIMM and consequent MVA depend on the production of computationally expensive training datasets, even when taking advantage of the statistical methods presented above. In this sense, our proposal is based on using a multi-output neural network as a proxy for DIM at monitoring times (which can be integrated to get MVA), incorporating one of the main ideas presented in \cite{Savine2020} (and also exploited, in a different financial context, in \cite{GomezCasanova2024}) for its training. Basically, they prove that, if you want to train a network to learn a quantity defined in terms of a conditional expectation on an initial state, you can use a training set made up of noisy estimations of such a quantity. Following this principle, we build training datasets for interest rate portfolios in which the IM-SIMM risk factors are encapsulated in interest rate model parameters and parsimonious representations of term structures (that serves as our network inputs, leading to a significant reduction in the number of model inputs), and whose labels (DIM targets) are highly noisy, simulated with a single MC sample. This strategy leads to two competitive advantages for the implementation in a real CCR engine. First, for the price of a single MC execution, we get a whole training dataset, and second, once the network has been fitted, it can predict DIM trajectories for a subset of interest rate model parameters (it parameterizes DIM trajectories). This is a major contribution since it allows, in a few minutes, to obtain precise DIM estimations at monitoring times of interest throughout the life of the desired portfolio, ready for use wherever needed.

The outline of the paper is as follows. In Section \ref{sec:Problem} we start by introducing the theoretical background for the DIM and MVA computation in the interest rate framework. Section 3 is devoted to the methodology followed in the elaboration of the training and validation datasets, as well as to the presentation of our proposal. In Section 4 we present numerical experiments to validate our methodology, essentially convergence and error analysis on a simple portfolio, consisting of a single swap; and the application to a more realistic portfolio. Finally, a summary of the work and conclusions are presented in Section \ref{sec:Conclusion}.

\section{Problem formulation} \label{sec:Problem}
 This section covers the introduction of relevant quantities when discussing CCR and collateralization, the basics of IM and the introduction of the standard methodology for its computation, as well as the theoretical background for the generation of prices and sensitivities of the proposed products and portfolios. Once these topics are presented, we consider its application in the calculation of DIM and MVA, which play an important role in producing the datasets used in network training and validation.

\subsection{Model assumptions}\label{sec:modelIR}
    From now on, we consider a frictionless and arbitrage-free continuous-time financial market with a finite time-horizon $T_f>0$. In addition, let $\left(\Omega,\mathcal{F},\Q\right)$ be the probabilistic space with the market filtration $\mathcal{F}$, given by $(\mathcal{F}_t)_{0\leq t\leq T_f}$, and the risk neutral probability measure $\Q$. Such measure is associated with the money market account $M_t$ as the numéraire, ensuring that all the attainable claims denominated by the numéraire are martingales under $\Q$, \cite{HARRISON1981}. 
    
    Given that we work with interest rate products, the short-rate dynamics we choose for the simulation of these products is of particular relevance. In particular, we restrict ourselves to work with Gaussian affine term-structure models since they provide closed-form formulas which allow the tractability of the computations, \cite{fabiomercurio}. Thus, we consider the Markovian state variables $x_t=\left(x_1(t),\dots, x_d(t)\right)\in \R^d$, and the smooth functions $\mu:[0, T_f]\times\R^d\to \R^d$, $\sigma: [0, T_f]\to\R^{d\times d}$.
    The state variables are assumed to verify the stochastic differential equation (SDE),
    \begin{equation}\label{eq:ir_sde}
        \text{d}x_t=\mu(t, x_t)\text{d}t+\sigma(t)\text{d}W_t,
    \end{equation}
    where $W_t=\left(W_1(t),\dots, W_d(t)\right)$ is a $d$-dimensional Brownian motion under $\Q$ and $\mu$ is an affine map in the $x_t$ components. Under these assumptions, we are able to define the short-rate $r_t$ as an affine map of the state variables $x_t$, leading to the advantage of being able to express the continuously compounded spot rate, commonly known as zero-rate, as an affine function of such a short-rate $r_t$. This relationship is exploited through the definition of zero-coupon bonds. Recall that a zero-coupon bond is a contract that guarantees its holder the payment of one unit of currency at time $T$, with no intermediate payments \cite{fabiomercurio}, and usually its value at time $t\leq T$ is denoted by $P(t, T)$. Under affine term-structure models the zero-coupon bond prices are exponential affine in $x_t$, i.e., they can be written in the form 
    \begin{equation*}
        P(t, T) = B(t, T)e^{-C(t, T)r(t)},
    \end{equation*}
    with $B$ and $C$ deterministic functions of time, which allows us to get the zero-rates via 
    \begin{equation*}
        Y(t, T) = -\dfrac{\log\left(P(t, T)\right)}{\tau\left(t, T\right)},
    \end{equation*}
    where $\tau(t, T)$ denotes the year fraction between $t$ and $T$. For the sake of simplicity, we adopt a single-curve framework in which the same short-rate is used for pricing and discounting.

\subsection{Pricing approach and instruments}
    Under the market assumptions established above, the general pricing formula of a financial derivative (or a portfolio of derivatives) under perfect collateralization with payoff $V(T)$ paid at time $T>t$ is given by
    \begin{equation}
        \label{eq:pricing}
        V(t) = \E^{\Q}\left[e^{-\int_t^Tr(u)\text{d}u}V(T)|\mathcal{F}_t\right].
    \end{equation}
    This general pricing formula can be clearly simplified in the case of the product we discuss, namely interest rate swaps (IRS). It should be noted that, if a portfolio of derivatives is considered, this can be reduced to the sum of the value of the individual contracts by means of \eqref{eq:pricing}. 

    \subsubsection{Interest rate swaps} \label{sec:irs_price}
        An IRS is a contract where one party agrees to pay the fixed cash flows that are equal to the interest at a predetermined, fixed rate on a notional amount (fixed leg) and where the other party pays a floating interests on the same notional amount (floating leg). We provide an informal approach to explain the swap pricing formula. Details of how to formally derive the expression from \eqref{eq:pricing}  can be found, for example, in \cite{oosterlee, MMMNF}.
        
        First, we introduce some notation relative to the payment schedules for the fixed and floating leg of the swap. Let be 
        \begin{equation}
        \begin{split}
            \T^P &= \{T_0^P, \dots, T_i^P, \dots, T_n^P \}, \\
            \T^R &= \{T_0^R, \dots, T_j^R, \dots, T_m^R\},
        \end{split}
        \end{equation}
        the fixed and floating schedules, respectively. Without loss of generality, we assume that $T_0^P=T_0^R$, $T_n^P=T_m^R$.
            
        Regarding the fixed leg, at every instant $T_i^{P} \in \T^P, i>0,$ it pays out the amount $NK\tau_P(T_{i-1}^P, T_i^P)$ corresponding to a fixed interest rate $K$, a nominal value $N$ and a year fraction convention $\tau_P$. In order to know the present value of the cash flows, it is necessary to apply the corresponding discounts. Thus, the discounted value at time $t$ is 
        \begin{equation}
            \label{eq:fixed_leg_value}
            PV_{\text{fixed}}(t; \T^P, K)=N \sum_{k = \eta_P(t)}^{n}\tau_P(T_{k-1}^P, T_k^P)P(t, T_k^{P})K,
        \end{equation}
        with $\eta_P(t) = \min\{{i\in\{1,\dots,n\}: T_i^P>t}\}$.
       
        The floating leg pays at every instant $T_j^{R} \in \T^R, j>0$ the amount $N\tau_R(T_{i-j}^R, T_j^R)L(T_{j-1}^{R}, T_j^{R})$, with $L(T_{j-1}^{R}, T_j^{R})$ the underlying spot float rate resetting at previous instant $T_{j-1}^{R} \in \T^R$. Discounting future cash flows, the present value of this leg at time $t$ is 
        \begin{equation}
            \label{eq:float_leg_value}
            PV_{\text{float}}(t; \T^R) = N\sum_{k=\eta_R(t)}^{m}\tau_R(T_{k-1}^R, T_k^R) P(t, T_j^{R}) F_k(t),
        \end{equation}
        where $\eta_R(t) = \min\{{j\in\{1,\dots,m\}: T_j^R>t}\}$ and $F_j(t)$ is the forward rate observed at time $t$, which can be written as
        \begin{equation*}
        F_j(t) =\begin{cases} \dfrac{1}{\tau_R(T_{j-1}^R, T_j^R)}\left(\dfrac{P(t, T_{j-1}^{R})}{P(t, T_j^{R})} - 1\right),& t<T_j^R,\\
        L(T_{j-1}^R, T_j^R),& T_{j-1}^R \leq t \leq T_{j}^R.
        \end{cases}     
        \end{equation*}        
        Thus, the price of the IRS at time $t$ is given by the difference between the present values of such legs \eqref{eq:fixed_leg_value}-\eqref{eq:float_leg_value}, i.e., 
        \begin{equation}
            \label{eq:swap_price}
            V_{\text{S}}(t; \T^P, \T^R, K, w) = wN\left[\sum_{k=\eta_R(t)}^{m}\tau_R(T_{k-1}^R, T_k^R) P(t, T_j^{R}) F_k(t) - \sum_{k = \eta_P(t)}^{n}\tau_P(T_{k-1}^P, T_k^P)P(t, T_k^{P})K \right],
        \end{equation}
        with $w=\in\{1, -1\}$ for payer or receiver swap, respectively. Typically, the swap is contractually configured to be at-the-money (ATM) at inception, i.e., the initial value of the swap is zero for both parties. The strike value for which the swap equals zero is called swap rate, $S(t; \T^P, \T^R)$, and is given by
         \begin{equation}
            \label{eq:swap_rate}
            S(t;\T^P, \T^R) = \dfrac{\sum_{k=\eta_R(t)}^{m}\tau_R(T_{k-1}^R, T_k^R) P(t, T_j^{R}) F_k(t)}{A(t; \T^P)}.
         \end{equation}
         where $A(t; \T^P) = \sum_{k = \eta_P(t)}^{n}\tau_P(T_{k-1}^P, T_k^P)P(t, T_k^{P})$ is known as the annuity factor, which represents the present value of a basis point of the swap.
         
        Note that the derived IRS pricing formula \eqref{eq:swap_price} is model-independent (due to its linear payoff), i.e., it does not require any interest rate model to compute its net present value. Furthermore, future swap prices can be computed with any model that  allows the valuation of zero-coupon bonds, justifying the adoption of affine term-structure models.

    \subsection{Collateral and Initial Margin}
        As discussed in the introduction, one of the major concerns in the financial world today is to be able to assess and mitigate the potential losses that an institution may incur in the event of a counterparty default. The most commonly used metrics for quantifying expected losses in the counterparty default are based on the concept of exposure, which represents the value that may be at risk. In particular, we focus on the expected positive exposure (EPE), which quantifies the expected losses in the event of a counterparty default; and on the potential future exposure (PFE), which indicates what the worst-case scenario for losses (with respect to a certain confidence level) would be at a future time, \cite{Gregory2020}.
        
        There are several mechanisms for reducing the institution's exposure to open positions with its counterparty, 
        including credit/debit value adjustments, netting strategies or collateralisation tools, such as VM and IM, being the later the one addressed here. IM can be seen as an amount that covers PFE for the expected time between the last VM exchange and the liquidation of positions on the default counterparty, \cite{BCBS}. The BCBS-IOSCO have jointly established rules for the calculation of IM. According to these rules, IM must be computed as a netting set $99\%$ confidence level PFE over a MPoR; and market participants have some freedom to develop internal models for the computation of said quantity, subject to validation and backtesting by regulators. Nevertheless, in the face of potential disputes between entities due to discrepancies in the calculated value, ISDA has introduced SIMM, a sensitive-based approach based on the risk profile of a position in terms of delta, vega and curvature sensitivities, defined across risk factors by tenor, expiry and asset class \cite{SIMM2013}. We adopt this model to calculate IM values required for our approach.

        Following \cite{isdameth}, this methodology distinguishes between four product classes: Interest Rates and Foreign Exchange (FX), Credit, Equity and Commodity, in such a way that every trade is assigned to an individual bucket. The total IM is given by the sum of IM-SIMM of each individual product class, i.e.,  
        \begin{equation*}
            \text{IM} = \text{SIMM}_{\text{RatesFX}} + \text{SIMM}_{\text{Credit}} + \text{SIMM}_{\text{Equity}} + \text{SIMM}_{\text{Commodity}}, 
        \end{equation*}
        with $\text{SIMM}_X$ the IM value of the product class $X$. In addition, every trade could be affected by a class-specific risk from the risk classes provided, namely: \textit{interest rate}, \textit{qualifying credit}, \textit{non-qualifying credit}, \textit{equity}, \textit{commodity} and \textit{foreign exchange}. Following this order, the SIMM for each product class $X$ is given by the variance-covariance formula
        \begin{equation*}
            \label{eq:IM_product}
            \text{SIMM}_{X} = \sqrt{\sum_r \text{IM}_r^2 + \sum_r\sum_{r\neq s}\psi_{rs}\text{IM}_r\text{IM}_s },
        \end{equation*}
        where $\text{IM}_r$ refers to the margin associated to each risk class and $\psi_{rs}$ is a correlation matrix between them given by ISDA. Lastly, the margin for each risk class is given by the sum of the delta, vega and curvature margins, i.e.,
        \begin{equation*}
            \text{IM}_r = \text{DeltaMargin}_r + \text{VegaMargin}_r + \text{CurvatureMargin}_r , 
        \end{equation*}
        where each component represents a PFE estimation with respect to the underlying risk factors. Details on how to calculate the  delta, vega and curvature margins can be found in \cite{isdameth} and our choice is described in Section \ref{sec:31}.
        
    \subsection{DIM and MVA}
        The use of different forms of collateral is associated with different funding costs. Since IM is a form of collateralisation that cannot be netted and must be posted into a segregated account, it is not rehypothecatable, leading to a considerable cost for the entities involved in the exchange. The risk associated with this funding cost is quantified by the MVA, introduced in \cite{Green2014}. Assessing MVA involves determining the price associated with holding IM throughout the lifetime of the transaction in question, which entails the need to calculate the IM at future times. Due to the stochastic nature of the underlying dynamics, the future IM value inherently represents a stochastic process \cite{Hoencamp2024}, whose valuation is performed by taking its conditional expectation. Thus, given the future IM, $\text{IM}(t),\,0\leq t\leq T$, DIM is understood as the expected IM collateral that has to be posted by the party at time $t$, and it is mathematically defined as 
        \begin{equation}
            \label{eq:DIM}
            \text{DIM}(t) = \E^{\Q}\left[e^{-\int_0^tr_u\text{d}u}\text{IM}(t)\left|\right.\mathcal{F}_0\right].
        \end{equation}        
        Once the DIM is known for each future instant of time, the MVA is computed by integrating the product of DIM by the spread of the collateral rate with respect to the risk-free rate until maturity, i.e., if we denote by $f(t)$ the funding spread at time $t>0$, 
        \begin{equation}
            \label{eq:MVA}
            \text{MVA} = \int_0^T f(s)\text{DIM}(s)\text{d}s.
        \end{equation}
        For the numerical experiments presented below, we assume a simplified expression for the funding spread given by the authors in \cite{Green2014} under certain funding strategy and contractual conditions. It leads to the following closed-form,
        \begin{equation}
            \label{eq:fsMVA}
            f(s) = \left(\left(1- R_B\right)\lambda_B(s)-s_I(s)\right)e^{-\int_{t_0}^s\left(\lambda_B(u)+\lambda_C(u)\right)\text{d}u},\quad s>t_0,
        \end{equation}
        where $R_B\in[0, 1]$ is the recovery rate of the seller B if its counterparty C defaults, $\lambda_B(s), \lambda_C(s)$ are the default intensities for each party at time $s$, and $s_I(s)$ is the spread on IM at time $s$.

\section{Methodology}
First, we define the kind of products/portfolios we focus on this work, mainly interest rate products without optionality, and the simplifications it entails in the IM-SIMM computation. Next, we show the nested MC scheme for computing DIM by brute force. This strategy is used to simulate our validation dataset. In addition, its presentation serves to highlight the problems it presents, mainly the computational cost due to the required nested simulations and re-evaluations. Then, we introduce the methodology we propose to approximate DIM using neural networks and the strategy followed to reduce the high training cost due to the dataset production.

\subsection{Simplifications IM-SIMM} \label{sec:31}
    As stated above, we only deal with interest rate products and/or portfolios, so the total IM is entirely represented by the IM-SIMM of the interest rate product class and is only affected by interest rate risk factors, mainly market interest rates (delta margin) and market implied volatilities (vega and curvature margin). Furthermore, since we work with products that are not subject to optionality, we can forget about implied volatilities (as the vega of such derivatives is zero) and work only with the interest rate risk factors. These considerations allow us to ensure that the IM of the described portfolios is entirely defined by the interest rate delta margin, i.e., $\text{IM} = \text{DeltaMargin}_{\text{IR}}$\footnote{It should be noted that, although we take these considerations into account in this work, the methodology described is of general applicability, and not only particular to the simplified case.}. 

    The interest rate risk factors are identified as the spine points of the market yield curve at the tenors $\T=\{2\text{W},1\text{M}, 3\text{M}, 6\text{M}, 1\text{Y}, 2\text{Y}, 3\text{Y}, 5\text{Y}, 10\text{Y}, 15\text{Y}, 20\text{Y}, 30\text{Y}\}$, and the sensitivity associated with each risk factor is given by the PV01 of the instrument \cite{isdameth}, i.e., if we denote by $s_k$ the sensitivity of the instrument $V_t$ associated with the tenor $\tau_k\in\T$, and $Y_k$ the yield value at that tenor, then $s_k=V_t(Y_k+1\text{bp}) - V_t(Y_k)$, where $1\text{bp}=0.01\%$. Once the $12$-dimensional sensitivity vector is calculated\footnote{Following the ISDA-SIMM requirements, if other tenors are used to compute the sensitivities, they have to be linearly allocated into the ISDA buckets.}, its entries are weighted by parameters provided by ISDA and then aggregated to provide the IM.

\subsection{Monte Carlo simulation of DIM} \label{sec:MCDIM}
    The simulation of collateralized exposure and XVA has a number of particularities associated with its application in the real world. We try to simplify these particularities as much as possible given the illustrative purpose of this article.
    We consider an interest rate derivative, or a portfolio of interest rate derivatives, subject to the assumptions stated in the previous section, and maturing at time $T>0$. We define a equispaced time grid $0=t_0<t_1<\dots<t_N=T$ with $N+1$ monitoring times and time step $h_t$, which ideally would be equal to the MPoR. We select the short-rate dynamics in line with the features set out in Section \ref{sec:modelIR} and assume its parameters are known via calibration procedure. We set as an initial condition the market state at time $t_0$, i.e., the spine points of the  yield curve at the ISDA-SIMM tenors, $Y^0 = \left(Y_1^0, Y_2^0,\dots, Y_{12}^0\right)$, which are obtained from the market rates for the same tenors, $R^0 = \left(R_1^0, R_2^0,\dots, R_{12}^0\right)$, via bootstrap procedure, \cite{Bianchetti2021}. Throughout the simulation, we need to interpolate discount values from the yield curve points. We apply a linear interpolation in yield between shock maturities given its simplicity, but any other method presented in \cite{Hagan2006} could be applied.

    Fixed a number $M$ of MC scenarios, for each scenario $1\leq j\leq M$ and monitoring time $t_i$,
    \begin{enumerate}
        \item We calculate the price of the instrument/portfolio, $V_j^i$, based on the forward market risk factors $Y_j^i=\left(Y_{1, j}^i,\dots, Y_{12, j}^i\right)$. These are directly dependent on the short-rate $r_j^i$, generated from the chosen dynamics and the value in the previous time $r_j^{i-1}$.
        \item Once the future price, $V_j^i$,  has been calculated, we compute the sensitivity of the instrument/portfolio with respect to the $k$-th forward risk factor, $s_{k, j}^i$, by shifting $Y_{k, j}^i$ one basis point and re-evaluating the product.
        \item The vector of sensitivities $\left(s_{1, j}^i,\dots, s_{12, j}^i\right)$ is then fed into the SIMM-IM calculation  engine to produce the IM value $\text{IM}_j^i$.
    \end{enumerate}
    Once we have simulated all the IM values for time $t_i$, we apply the MC estimator to \eqref{eq:DIM} for approximating the DIM value at time $t_i$, $\text{DIM}_i$, i.e., 
    \begin{equation*}
        \text{DIM}_i = \dfrac{1}{M}\sum_{j=1}^{M}\text{IM}_j^ie^{-\sum_{l\leq i}r_j^ih_t }.
    \end{equation*}
    A summary of the DIM calculation for a given monitoring time can be found in Algorithm \ref{alg:MC-DIM}. It should be noted that the error in the MC estimate is approximately normally distributed with mean $0$ and standard deviation $\sigma_{i}/\sqrt{M}$, \cite{glasserman}, where $\sigma_i$ is the standard deviation of DIM at time $t_i$, portfolio dependent and usually unknown. Thus, the method presents a square-root convergence rate and adding one decimal place of precision requires $100$ times as many paths considered, making accurate DIM estimates very costly to obtain.

    Finally, once the DIM has been calculated for all monitoring times, we can approximate the MVA of the instrument/portfolio. The standard approach we follow consists of applying a simple quadrature rule to \eqref{eq:MVA},
    \begin{equation} \label{eq:MVAdisc}
        \text{MVA}\approx \sum_{i=1}^{N}f_i\text{DIM}_ih_t,
    \end{equation}
    where $f_i$ is the funding spread at time $t_i$.
    
    \begin{algorithm}[thb]
        \caption{Nested MC simulation of DIM}
        \label{alg:MC-DIM}
        \begin{algorithmic}[1]
            \Require Simulation times $0=t_0<t_1<\dots<t_N=T$ with $h_t$ time step. Yield curve points at the ISDA-SIMM tenors for time $t_0$, $\{Y^0_k\}_{k=1}^{12}$. Model parameters $\Theta$. Number of MC paths, $M$. Samples of the standard normal distribution i.i.d., $(W)_j^i,\,j=0\dots,M-1,\,i=0\dots N-1$.
            
            \Ensure Compute $\text{DIM}_i$, the dynamical IM at time $t_i$. 
            \State $x^0 \gets Y^0$ 
            \Comment{Get the initial state variables $x_0$ from the initial yield curve $Y^0$}.
            \For{$i=1$, $i<N+1$, $i++$}
                \For{$j=0$, $j<M$, $j++$}
                    \State $x_j^i \gets \left(x_{j}^{i-1}, W_j^{i-1}, \Theta\right)$ \Comment{Get state variable $x_j^i$ from the previous state via IR dynamics.}
                    \State $\{Y_{k,j}^i\}_{k=1}^{12} \gets \left(x_j^i, \Theta\right)$ \Comment{Obtain yield curve points from the current state vars. and model.}
                    \State $V_j^i \gets \{Y_{k,j}^i\}_{k=1}^{12}$ \Comment{Get the market to future value.} 
                    \For{$k=1$, $k<13$,$k++$}
                        \State $Y_{k, j}^i \gets Y_{k, j}^i + 1\text{bp}$ \Comment{Apply shock to the $k$-th yield.}
                        \State $V_{k, j}^i \gets \{Y_{k,j}^i\}_{k=1}^{12}$ \Comment{Re-evaluate instrument.}
                        \State $S_{k, j}^i \gets V_{k, j}^i - V_i^j$ \Comment{Compute sensitivity w.r.t. $k$-th yield.}
                        \State $Y_{k, j}^i \gets Y_{k, j}^i - 1\text{bp}$ \Comment{Set $k$-th yield to its true value.}
                    \EndFor
                    \State $\text{IM}_j^i\gets \{S_{k, j}^i\}_{k=1}^{12}$ \Comment{IM is compute from the sensitivities via SIMM}
                    \State $\text{IM}_j^i\gets \text{IM}_j^ie^{-\sum_{\hat{k}=0}^ir^j_{\hat{k}}(t_{i+1}-t_{i})}$ \Comment{Apply discount to $t_0$.}
                \EndFor
                \State $\text{DIM}_i\gets \dfrac{1}{M}\sum_{j=0}^M\text{IM}_j^i$ \Comment{Approximate DIM from its MC estimator.}
            \EndFor
        \end{algorithmic}
    \end{algorithm}

 \subsection{DIM learning} \label{sec:DIMlearning}
    The nested MC methodology presented in Section \ref{sec:MCDIM} highlights the difficulties of being applied in realistic situations. At each time step and MC scenario, as many portfolio re-evaluations are performed as many risk factors are considered. Given the possibility that a portfolio may be composed of hundred of products and complicated exotic transactions, such simulation becomes impractical due to the computational cost and time required. Moreover, the situation becomes even worse if one wants to compute such values for different initial market states.

    These circumstances lead to considering the use of machine learning and/or deep learning techniques to approximate the function that defines, in this case, the DIM values. From now on, we define the deterministic DIM function as $\mathbf{F}:\Omega\subset \mathbb{R}^d \to \mathbb{R}^N$, the time discretized version of the equation \eqref{eq:DIM}, where $\Omega$ is the $d$-dimensional feature space of market state variables at time $t_0$, and $N$ is the number of DIM monitoring times. This multi-output structure is one of the design choices we made for our model approximation. It entails removing time as an input variable (leaving only the initial market state variables as inputs to the model) and treating the output as the set of temporal reference points at which the DIM is intended to be known.

    Such definition poses no issue due to the point-wise nature of conditional expectations. Let be $\mathbf{X}$ the market state variable, and $\mathbf{IM} = \left(\text{IM}_0,\dots, \text{IM}_N\right)$ the vector random variable, understood as the discretization of the IM stochastic process defined above. From \eqref{eq:DIM} it follows that 
    \begin{equation}
    \begin{aligned}
        \mathbf{DIM}=\left(\text{DIM}_0,\dots, \text{DIM}_N\right) &= \mathbb{E}^{\mathbb{Q}}\left[\left(\text{IM}_0,\dots,\text{IM}_N\right)\Bigl|\mathbf{X}\right] = \Bigl(\mathbb{E}^{\mathbb{Q}}\left[\text{IM}_0\bigl|\mathbf{X}\right],\dots,\mathbb{E}^{\mathbb{Q}}\left[\text{IM}_N\bigl|\mathbf{X}\right] \Bigr)\\
    \end{aligned}
    \end{equation}
    In addition, if we assume that $\mathbf{X}$ is the complete Markov state at time $t_0$, there exist a deterministic, but unknown, function $\text{F}_i$ of the state. Thus, our multi-output function $\mathbf{F}$ is defined as
    \begin{equation}
        \mathbf{F}(\mathbf{X}) = \left(\text{F}_0\left(\mathbf{X}\right),\dots, \text{F}_N\left(\mathbf{X}\right)\right) = \Bigl(\mathbb{E}^{\mathbb{Q}}\left[\text{IM}_0\bigl|\mathbf{X}\right],\dots,\mathbb{E}^{\mathbb{Q}}\left[\text{IM}_N\bigl|\mathbf{X}\right] \Bigr).
    \end{equation}

\begin{figure}
        \centering
        \includegraphics{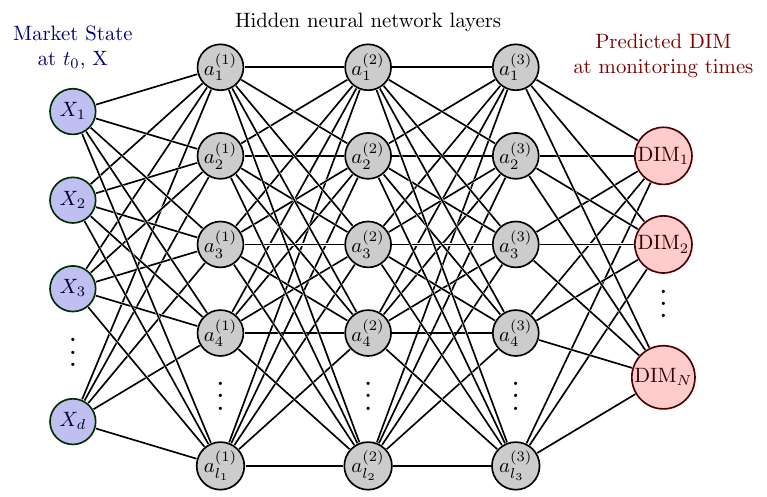}
        \caption{Schematic representation of the multi-output fully connected neural network used as a proxy $\mathbf{\hat{F}}$ for DIM values at monitoring times. Each connecting edge represents an updatable weight. Each vertex in the hidden layers represents the weighted sum of its inputs (values of the previous layer) and a non-linear function acting over such a sum.}
        \label{fig:NN}
    \end{figure}

    The idea of ML/DL applied to our context is to approximate our defined function $\mathbf{F}$ through the mapping $\mathbf{\hat{F}}\left(x; \boldsymbol{\omega}\right)$, with the weights $\boldsymbol{\omega}$ that results in the best approximation. The most common choice for $\mathbf{\hat{F}}$ in this kind of numerical approximation tasks are feed forward networks, consisting of the composition of several layers, where each layer is also a composition of affine maps and non-linear transformations (see Figure \ref{fig:NN}). The coefficients of the affine maps are the above-mentioned weights that need to be adjusted to fulfill the approximation requirements. Such process is called training and can be performed in several ways.

    \subsubsection{Classical supervised learning approach}
        The first approach that one might think of is to train our DIM network by supervised learning. In this training strategy, the weights are updated from a training set consisting of samples drawn from the initial market state variable, $\mathbf{X}^{(k)}$, and their corresponding vector DIM values, $\mathbf{DIM}^{(k)}$. They are simulated via Algorithm \ref{alg:MC-DIM} and can be consider ground truth values if the number of Monte Carlo paths $M$ is large enough.

        In each training iteration, the network is evaluated in a market state sample $\mathbf{X}^{(k)}$ and the result is compared with the respective label $\mathbf{DIM}^{(k)}$ by means of a cost function that measures how far the prediction is from the target.
        Thus, if we assume that $\Psi(\cdot, \cdot)$ is a chosen metric, we can define the cost function as
        \begin{equation*}
            \J: \boldsymbol{\omega}\in  \mathbb{R}^{\text{l}(\boldsymbol{\omega})} \to \J(\boldsymbol{\omega})= \Psi\left(\mathbf{\hat{F}}(\,\,\cdot\,\,; \boldsymbol{\omega}), \mathbf{F}(\,\cdot\,)\right) \in \mathbb{R},
        \end{equation*}
        where $\text{l}(\boldsymbol{\omega})$ is the dimension of the parameter space. The goal of the training is to find $\boldsymbol{\omega}^*\in \mathbb{R}^{\text{l}(\boldsymbol{\omega})}$ such that
        \begin{equation*}
            \boldsymbol{\omega}^* = \arg\min_{ \boldsymbol{\omega}\in \mathbb{R}^{\text{l}(\boldsymbol{\omega})}}\J(\boldsymbol{\omega}).
        \end{equation*}
        In our case, we decide to work with the squared $L^2$ norm, so we get
        \begin{equation}
        \begin{aligned}
            \Psi\left(\mathbf{\hat{F}}(\mathbf{X}^{(k)}; \boldsymbol{\omega}), \mathbf{F}(\mathbf{X}^{(k)})\right) &= \Psi\left(\mathbf{\hat{F}}(\mathbf{X}^{(k)}; \boldsymbol{\omega}), \mathbf{DIM}^{(k)}\right) \\ &
            = \bigl|\bigl|\mathbf{\hat{F}}(\mathbf{X}^{(k)}; \boldsymbol{\omega})-\mathbf{DIM}^{(k)}\bigr|\bigr|^2_2 \\ 
            &= \sum_{i=0}^{N+1}\left(\hat{\text{F}}_i(\mathbf{X}^{(k)}; \boldsymbol{\omega}) - \text{DIM}_i^{(k)}\right)^2.
        \end{aligned}
        \end{equation}
        In practice, the process of updating weights is done after evaluating the approximator in a given number of samples (batch size), then the real loss estimation used is the mean of the squared errors (MSE).
    
        However, this kind of strategy is in principle not valid. The computational cost of producing the training dataset with ground truth DIM samples is the same as the cost of using brute force to obtain the DIM value, which does not provide the necessary speed and flexibility to be able to include this type of approach in a real CCR engine.

    \subsubsection{Supervised learning with sampled labels} \label{subsec:LSMLearning}
    
        In order to improve the situation described above, what we can do is to take advantage of one of the main ideas introduced by the authors in \cite{Savine2020}. Thus, they proved that it is possible to approximate functions defined in terms of conditional expectations by means of datasets consisting of unbiased noisy labels, which can even be generated with a single MC path per initial state. Based on this idea, we can write  
        \begin{equation}
        \begin{aligned}
            \mathbf{DIM} = \mathbb{E}^{\mathbb{Q}}\left[\mathbf{IM}|\mathbf{X}\right] &= \min_{\mathbf{F}\in L^2\left(\mathbf{X}\right)}\bigl|\bigl|\mathbf{F}(\mathbf{X})-\mathbf{IM}\bigr|\bigr|_2^2 \\
            &\approx \min_{\boldsymbol{\omega}\in\mathbb{R}^{\text{l}(\boldsymbol{\omega})}} \bigl|\bigl|\hat{\mathbf{F}}(\mathbf{X}, \boldsymbol{\omega})-\mathbf{IM}\bigr|\bigr|_2^2 \\ 
            &\approx \hat{\mathbf{F}}\left(\mathbf{X}, \min_{\boldsymbol{\omega}\in\mathbb{R}^{\text{l}(\boldsymbol{\omega})}}\text{MSE}\right),
        \end{aligned}
        \end{equation}
        by the universal approximation theorem and the assumption of an IID dataset sampled from the correct distribution. More details can be found in \cite{Savine2020} (Appendix 1).

\begin{figure}
            \centering
            \includegraphics[width=0.75\linewidth]{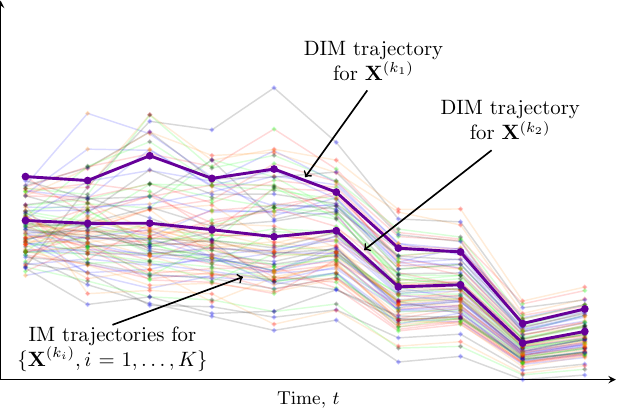}
            \caption{Schematic representation of the strategy explained in Section \ref{subsec:LSMLearning}. The graphs in faded colors represent the set of IM samples (noisy DIM estimations), each from a different initial market state, used as training dataset (sampled labels). The highlighted lines represent the approximations of the DIM trajectories for two initial states obtained by the DIM network trained with these sampled labels. As mentioned above, we work with a discrete version of these trajectories, represented in the figure by the points at each monitoring time.}
            \label{fig:imdimm}
        \end{figure}

        By application of this result, we no longer have to produce a dataset with ground truth DIM values. It will only be necessary to compute noisy estimations via Algorithm \ref{alg:MC-DIM}. Following this philosophy, we compute each dataset label taking a single MC path ($M=1$). This represents a major advantage since, for the computational cost of one truly converged Monte Carlo DIM execution, we can now produce an entire dataset to train our DIM network. This allows us to overcome the difficulties presented before and justify its use in real applications. This idea is schematically shown in Figure \ref{fig:imdimm}.

    \label{sec:Methodology}
\section{Numerical experiments}
In this section, we present numerical experiments concerning the convergence of the algorithm as a function of the number of samples used in training, as well as a study on the distribution of the error incurred by the trained network in relation to its input data. Finally, we validate our approach by computing the DIM and the MVA in a practical situation.

\subsection{Experiment configuration}
    We choose to assess the experiments in two different interest rate models, namely the Vasicek, \cite{Vasicek1977}, and the Hull-White, \cite{HullWhite1990}, models. Both follow the general structure given in equation \eqref{eq:ir_sde} and can be written as 
    \begin{equation*}
        \begin{aligned}
            \text{d}x(t) &= -ax(t)\text{d}t + \sigma\text{d}W_t,\\
            r(t) &= x(t) +  \theta(t),        
        \end{aligned}
    \end{equation*}
    where $a>0, \sigma>0$ are the mean reversion speed and the volatility of the process. $\theta(t)$ can be interpreted as the long term average rate, but it plays a different role in the considered models. In Vasicek, $\theta(t)=\theta$ is taken as a constant parameter and, to completely determine the stochastic equation, we need to add the initial short-rate $r_0$. Under this setting, the market state at time $t_0$ is completely determined by these parameters, i.e., they are inputs in our methodology, see Table \ref{tab:ParamsSetting}. It should be noted that both the initial and the time-future yield curve points scenarios, i.e., the market risk factor used to compute the ISDA SIMM, are fully determined under these parameters, since the model defines the present and the future realizations of the yield term structure, obviously conditional on the current state.  
    
    One of the main drawbacks of the Vasicek model is that it cannot be fitted to the yield term structure observed in the market at time $t_0$. Because of that, we want to introduce a more realistic example of application with the Hull-White model, which does allow for fitting the current yield curve. In such a model, $\theta(t)$ is a function that depends on the $t_0$ market instantaneous forward rate, i.e., which depends on the aforementioned term structure at time $t_0$, see, e.g., \cite{fabiomercurio}, but the initial condition is always the same, $x_0=0$. 
    
    In order to summarise the relevant information in the yield curve, we use a parsimonious model for the latter. Thus, it is carried out by means of the Nelson-Siegel parametrisation, \cite{NelsonSiegel1987}, which is defined from four parameters: $\beta_0, \beta_1, \beta_2$ and $\lambda>0$. In this work we set $\lambda=1.37$, a common choice in the industry, \cite{Annaert2000}. Under these premises, the market state at time $t_0$ is now determined by the Hull-White model parameters $a, \sigma$, jointly with the free Nelson-Siegel parameters $\beta_0, \beta_1, \beta_2$, see Table \ref{tab:ParamsSetting}.
    
    For testing purposes, we consider a portfolio made up with a single product, a $1$ year forward, $5$ years expiry interest rate swap with quarterly-annual float and semi-annual fixed leg schedules. We are interested in present results not only for the ATM swap, but also for in-the-money (ITM) and out-the-money (OTM) versions, so we add to the previously considered market state variables the spread, $\delta$, between the fixed ATM rate (completely determined by the yield term structure at time $t_0$) and the one chosen.

    \begin{table}[t]
        \centering
        \begin{tabularx}{0.45\linewidth}{ X X X X X X}
            \multicolumn{6}{c}{Vasicek setting}\\
            \toprule
             \multicolumn{1}{c}{$\bold{X}$}  &  \multicolumn{1}{c}{$a$} & \multicolumn{1}{c}{$\sigma$}& \multicolumn{1}{c}{$\theta$} & \multicolumn{1}{c}{$r_0$} & \multicolumn{1}{c}{$\delta$}\\
             \midrule
             \multicolumn{1}{c}{$\text{min}(\bold{X})$}  &  \multicolumn{1}{c}{$1\%$} & \multicolumn{1}{c}{$0.5\%$} & \multicolumn{1}{c}{$0.1$} & \multicolumn{1}{c}{$-5\%$} & \multicolumn{1}{c}{$-0.1\%$} \\
            \multicolumn{1}{c}{$\text{max}(\bold{X})$}  &  \multicolumn{1}{c}{$10\%$} & \multicolumn{1}{c}{$2.5\%$} & \multicolumn{1}{c}{$5\%$} & \multicolumn{1}{c}{$5\%$} & \multicolumn{1}{c}{$0.1\%$} \\
            \bottomrule
        \end{tabularx}
    
        \bigskip
    
        \begin{tabularx}{0.5\linewidth}{ X X X X X X X}
            \multicolumn{7}{c}{Hull-White setting}\\
            \toprule
             \multicolumn{1}{c}{$\bold{X}$}  &  \multicolumn{1}{c}{$a$} & \multicolumn{1}{c}{$\sigma$}& \multicolumn{1}{c}{$\beta_0$} & \multicolumn{1}{c}{$\beta_1$} & \multicolumn{1}{c}{$\beta_2$}& \multicolumn{1}{c}{$\delta$}\\
             \midrule
             \multicolumn{1}{c}{$\text{min}(\bold{X})$}  &  \multicolumn{1}{c}{$1\%$} & \multicolumn{1}{c}{$0.05\%$}& \multicolumn{1}{c}{$-0.5\%$} & \multicolumn{1}{c}{$0\%$} & \multicolumn{1}{c}{$0\%$}& \multicolumn{1}{c}{$-0.1\%$}\\
            \multicolumn{1}{c}{$\text{max}(\bold{X})$}  &  \multicolumn{1}{c}{$5\%$} & \multicolumn{1}{c}{$1.5\%$}& \multicolumn{1}{c}{$5\%$} & \multicolumn{1}{c}{$1\%$} & \multicolumn{1}{c}{$1\%$}& \multicolumn{1}{c}{$0.1\%$}\\
            \bottomrule
        \end{tabularx}
        \caption{Lower and upper bounds for the market state variables in each considered model.}
        \label{tab:ParamsSetting}
    \end{table}
    
    To carry out the experiments, we need to generate both the training set and the validation set for each proposed setting. We take $N=160$ equispaced monitoring times between the settle time of the swap, $t_0=0$, and its maturity, $T_f=6$. To generate both datasets, we follow the methodology explained in Section \ref{sec:MCDIM}. We sample $K$ vectors of initial market state values from the given setting (Vasicek or Hull-White) via Latin Hypercube Sampling (LHS), \cite{glasserman}, with the upper and lower bounds specified in Table \ref{tab:ParamsSetting}. The main difference appears in the hyperparameters of the simulation procedure. On the one hand, the training dataset is developed taking $K_T=2^{22}$ market state vectors. For each of them, we get, following the Algorithm \ref{alg:MC-DIM}, a noisy vector of DIM estimations by setting $M_T=1$ MC paths. On the other hand, the validation dataset is made up of $K_V=2^{9}$ market state vectors and, each corresponding DIM trajectory is again computed with the aforementioned Algorithm \ref{alg:MC-DIM} taking $M_V=2^{20}$ MC paths. We consider such DIM vectors as the ground truth estimations for the quantity. Nevertheless, recall that, by the nature of the MC algorithm, we have some error in our validation sets. Furthermore, the error varies depending on both the model and the chosen monitoring time, so we consider as a tolerance the maximum MC error deviation estimated in each model, namely, $4.24\times10^{-4}$ for Vasicek and $2.12\times10^{-4}$ for Hull-White. 

    The network architecture used as a proxy follows the guidelines set out in Section \ref{sec:DIMlearning}. Essentially, it is a multi-output fully connected feed forward network with as many outputs as monitoring times. The hyperparameters configuration is presented in Table \ref{tab:NNsttruct}. Only note that, except for the number of inputs, we keep the same hyperparameters in both settings. A pre-processing layer is added, which normalises the input data to values between $0$ and $1$. Regarding the training stage, we choose to use the most common in literature stochastic gradient-base optimizer, Adam, \cite{Adam2014}, with its default configuration. The train is performed in batches of $4096$ samples. We apply a learning rate policy as the training progresses. From an initial learning rate of $10^{-3}$, we halve it if the mean square test error does not show improvement within $50$ iterations, until it reaches the value of $10^{-6}$. We set the number of epochs as $2000$ and an early-stopping criterion is added to avoid wasting time once the threshold accuracy has been reached. All the experiments are executed on a system AMD EPYC 7763 64-core processor, 32 GB of RAM, equipped with a NVIDIA A100 GPU. The codes were implemented in Python 3.8 and Tensorflow v$2.9$.
    
    Lastly, to check how well the proposed methodology works, we measure the impact it has on the MVA calculation \eqref{eq:MVA}. To this end, we consider the counterparty risk parameters given in \cite{Green2014} to account for the funding spread \eqref{eq:fsMVA}: $\lambda_C=0$, $\lambda_B=1.67\times 10^{-2}$, $R_B=0.4$ and $s_I=0$; and we compute the MVA for each ground truth DIM path by means of \eqref{eq:MVAdisc}.
    
    \begin{table}[t]
        \centering
        \begin{tabularx}{0.63\linewidth}{X X X}
            \toprule
            &\multicolumn{1}{c}{Vasicek setting} & \multicolumn{1}{c}{Hull-White setting}\\
            \toprule
            \toprule
            Num. inputs & \multicolumn{1}{c}{$5$} & \multicolumn{1}{c}{$6$} \\
            \midrule
            Num. outputs & \multicolumn{2}{c}{$160$} \\
            \midrule
            Num. hidden layers & \multicolumn{2}{c}{$3$}\\
            \midrule
            Num. units per layer & \multicolumn{2}{c}{$256$} \\
            \midrule
            Activation function & \multicolumn{2}{c}{$\sigma(x)=\max\left(0, x\right)$ (ReLU)} \\
            \midrule
            Weight initialization & \multicolumn{2}{c}{Glorot uniform, \cite{Glorot2010}}\\
            \bottomrule
        \end{tabularx}
        \caption{Feed forward network hyperparameters for each setting.}
        \label{tab:NNsttruct}
    \end{table}
    
\subsection{Convergence in terms of training samples}
    The first test we present aims to study training behavior as a function of the number of employed sampled labels. It is expected that, as the training set grows in number of samples, the accuracy of the approximation will improve, but by how much? The idyllic situation would be to get the same convergence ratio as given by MC algorithm. This would mean that, for the price of a single MC to generate the training set, plus the training cost, we would have a fitted network that give us the numerical approximation of DIM at monitoring times, with the same accuracy, for a whole space of initial market conditions. To this end, from the total training set of sampled labels, we select shuffled subsets with a number of samples ranging from $512$ to the total, $2^{22}$, in powers of two. With each subset, we train $20$ neural networks with the guidelines set out above. We compare their performance by looking at two metrics against the validation set, the square root of the MSE (RMSE) and the MVA computation (with the risk parameters given before). 
    
\begin{figure}[ht] 
        \centering
        \begin{subfigure}[b]{0.49\textwidth}
            \centering
            \includegraphics[page=1, width=\textwidth]{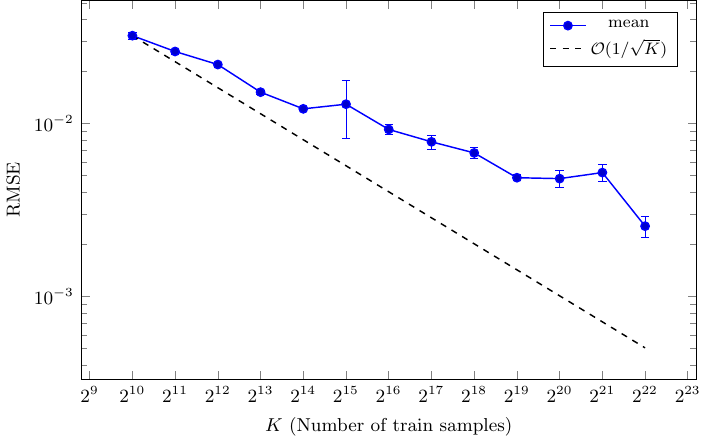}
            \caption{Vasicek RMSE}
        \end{subfigure}
        \hfill
        \begin{subfigure}[b]{0.49\textwidth}
            \includegraphics[page=2, width=\textwidth]{Plots/vasicek_numsamples.pdf}
            \caption{Vasicek MVA absolute error}
        \end{subfigure}
        \hfill
        \vspace{0.5cm}
        \begin{subfigure}[b]{0.49\textwidth}
            \centering
            \includegraphics[page=1, width=\textwidth]{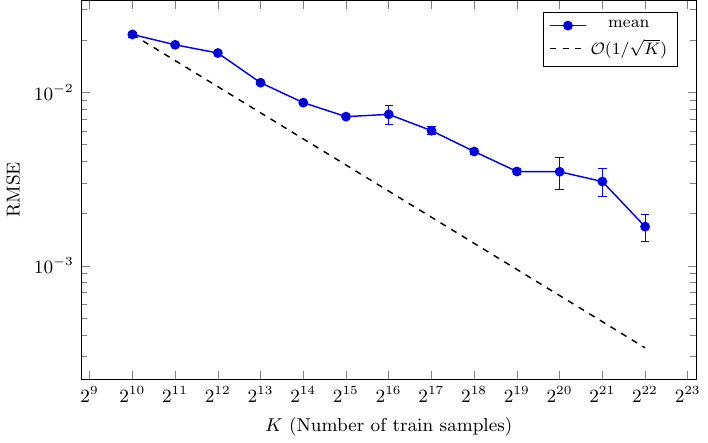}
            \caption{Hull-White RMSE}
        \end{subfigure}
        \hfill
        \begin{subfigure}[b]{0.49\textwidth}
            \includegraphics[page=2, width=\textwidth]{Plots/hullwhite_numsamples.pdf}
            \caption{Hull-White MVA absolute error}
        \end{subfigure}
        \caption{Convergence for each dataset size and model setting. For each one, the mean and $95\%$ confidence interval error is computed with $20$ training trials. Square-root convergence rate is shown for comparative purposes.}
        \label{fig:testNS}
    \end{figure}

    In Figure \ref{fig:testNS}, we present the mean value of the two metrics for each interest rate setting and training subset, as well as a $95\%$ confidence interval estimated with a $t-$distribution. The square-root convergence rate is also plotted for comparative purposes. In practically all cases, the variability of the training is very small, especially in terms of calculating MVA. The only case that we can highlight in terms of variability is the Vasicek setting with $K_T=2^{15}$, which was produced by the divergence of one of his trials. As expected, as the number of samples in the training dataset increases, there is an error reduction, but the idyllic behavior is not met, the error reduction looses the MC convergence order. Even so, we find the result very satisfactory. The loss of order is not very pronounced, and we should remember that the trained networks parameterize DIM for a whole space of initial conditions. Furthermore, no significant differences are observed between settings.
    
\subsection{DIM error in terms of the initial state variables}
    Having shown the overall error (in terms of RMSE and MVA) with respect to the number of training samples, we are now interested in studying how the difference between ground truth and predicted DIM is distributed across the initial state variables. For that, we analyze the results obtained by the networks trained with the largest dataset size in the previous test. In particular, we focus on the differences given in the monitoring time which presents a higher DIM variance, since larger values are expected. In our simple portfolio case, where the only product is an interest rate swap, we know that the instant of highest DIM variance it just before the first cashflow exchange, so we set the monitoring time at $t_{\gamma} = 1.75$ for both settings. 

\begin{figure}[ht]
        \centering
        \begin{subfigure}[b]{0.32\textwidth}
            \centering
            \includegraphics[page=1, width=\textwidth]{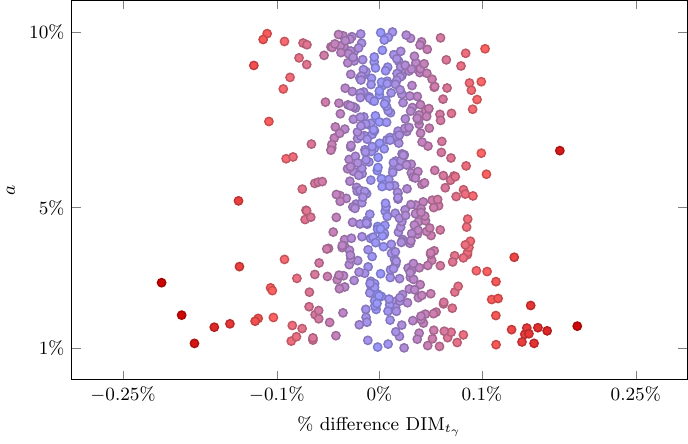}
        \end{subfigure}
        \begin{subfigure}[b]{0.32\textwidth}
            \centering
            \includegraphics[page=2, width=\textwidth]{Plots/vasicek_params_errors.pdf}
        \end{subfigure}
        \begin{subfigure}[b]{0.32\textwidth}
            \centering
            \includegraphics[page=3, width=\textwidth]{Plots/vasicek_params_errors.pdf}
        \end{subfigure}
        \vspace{0.5cm}
        \begin{subfigure}[b]{0.32\textwidth}
            \centering
            \includegraphics[page=4, width=\textwidth]{Plots/vasicek_params_errors.pdf}
        \end{subfigure}
        \begin{subfigure}[b]{0.32\textwidth}
            \centering
            \includegraphics[page=5, width=\textwidth]{Plots/vasicek_params_errors.pdf}
        \end{subfigure}
        \caption{Mean differences for DIM, at monitoring time $t_{\gamma}=1.75$, between ground truth and predictions in the Vasicek setting, in terms of the input variables $a$ (top left), $\sigma$ (top center), $\theta$ (top right), $r_0$ (bottom left) and $\delta$ (bottom right).}
        \label{fig:difV}
    \end{figure}

\begin{figure}[ht]
        \centering
        \begin{subfigure}[b]{0.32\textwidth}
            \centering
            \includegraphics[page=1, width=\textwidth]{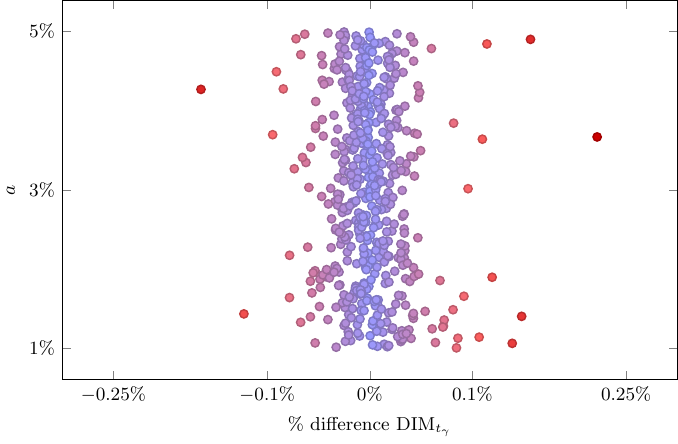}
        \end{subfigure}
        \begin{subfigure}[b]{0.32\textwidth}
            \centering
            \includegraphics[page=2, width=\textwidth]{Plots/hullwhite_params_errors1.pdf}
        \end{subfigure}
        \begin{subfigure}[b]{0.32\textwidth}
            \centering
            \includegraphics[page=3, width=\textwidth]{Plots/hullwhite_params_errors1.pdf}
        \end{subfigure}
        \vspace{0.5cm}
        \begin{subfigure}[b]{0.32\textwidth}
            \centering
            \includegraphics[page=1, width=\textwidth]{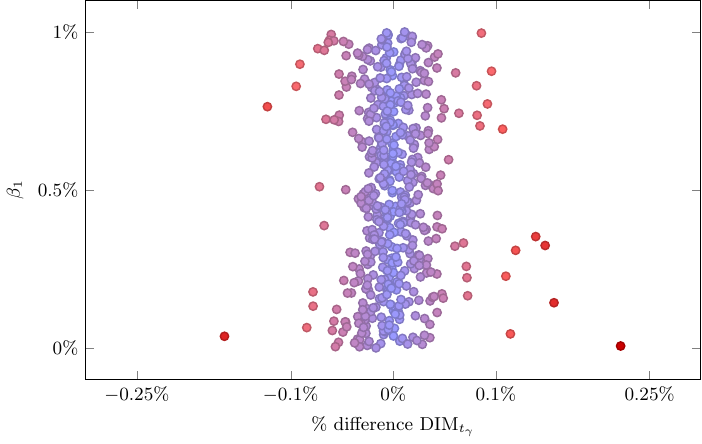}
        \end{subfigure}
        \begin{subfigure}[b]{0.32\textwidth}
            \centering
            \includegraphics[page=2, width=\textwidth]{Plots/hullwhite_params_errors2.pdf}
        \end{subfigure}
        \begin{subfigure}[b]{0.32\textwidth}
            \centering
            \includegraphics[page=3, width=\textwidth]{Plots/hullwhite_params_errors2.pdf}
        \end{subfigure}
        \caption{Mean differences for DIM, at monitoring time $t_{\gamma}=1.75$, between ground truth and predictions in the Hull-White setting, in terms of the input variables $a$ (top left), $\sigma$ (top center), $\beta_0$ (top right), $\beta_1$ (bottom left) and $\beta_2$ (bottom center) and $\delta$ (bottom right).}
        \label{fig:difHW}
    \end{figure}

    In Figure \ref{fig:difV} and Figure \ref{fig:difHW} we present the differences obtained for the Vasicek and Hull-White, respectively. Both results are presented in terms of the mean difference calculated from the $20$ trained networks. In the light of the outcome, we extract the following conclusions:
    \begin{itemize}
        \item The differences in both settings are centered around zero, i.e., no bias is observed, and bounded by $0.25\%$. We see more dispersion in the Vasicek case, which is to be expected given that its estimated variance is twice the variance calculated in Hull-White.
        \item In general, the largest differences are observed in the boundary values of each market variable. This is a common pattern in supervised training algorithms since the number of samples in these regions is less dense. 
        \item Parameter values that increase the variance generate larger differences. This is most clearly seen in the Vasicek case, where the combination given by small mean reversion speeds, $a$, with large model volatilities, $\sigma$, leads to higher DIM variances and higher prediction differences. In Hull-White this pattern is less pronounced, but the positive correlation between model volatility and prediction differences is evident.
    \end{itemize}

    \subsection{DIM and MVA error in terms of the interest rate model volatility}
    
    As we have seen, the greatest source of error in the prediction comes from extreme initial state values, especially of the parameters that control the volatility of the interest rate model used. For this reason, we generate ad-hoc samples of DIM trajectories to more comprehensively evaluate such cases. In the Vasicek setting, we take $\theta=3\%$, $r_0=1\%$, $\delta=0\%$, and we analyze the performance of the trained networks for combinations of $a=1\%,\,5\%,\,10\%$ and $\sigma=0.5\%,\,1\%,\,2.5\%$. The same strategy is followed with Hull-White, where we set $\beta_0=1\%$, $\beta_1=\beta_2=0.5\%$, $\delta=0\%$ and combinations of $a=1\%,\,2.5\%,\,5\%$, $\sigma=0.05\%,\,0.75\%,\,1.5\%$ are taken. Both cases present the same behavior, their intrinsic volatility increase with small values of $a$ and with large values of $\sigma$. Note that the simulation of the DIM values in the described cases keeps all the properties specified at the beginning of the section.
    
    \begin{table}[ht]
        \centering
        \begin{tabular}{|m{1.5cm} m{3cm}|c c c|}
            \hline
            \multirow{2}{*}{$a$} & \multirow{2}{*}{Target} & \multicolumn{3}{c|}{$\sigma$} \\ \cline{3-5} 
             &  & $0.5\%$ & $1\%$ & $2.5\%$ \\ \hline \hline
            \multirow{3}{*}{$1\%$} & DIM$(t=t_{\gamma})$ & $3.67\times10^{-4}$ & $3.58\times10^{-4}$ & $2.95\times10^{-3}$ \\ 
             & DIM$(t=2t_{\gamma})$ & $1.14\times10^{-5}$ & $1.76\times10^{-4}$ & $3.55\times10^{-3}$ \\ 
             & MVA & $1.08\times10^{-3}$ & $3.63\times10^{-4}$ & $2.23\times10^{-3}$ \\ \hline
            \multirow{3}{*}{$5\%$} & DIM$(t=t_{\gamma})$ & $1.07\times10^{-4}$ & $1.35\times10^{-4}$ & $4.66\times10^{-4}$ \\ 
             & DIM$(t=2t_{\gamma})$ & $1.80\times10^{-4}$ & $2.55\times10^{-4}$ & $8.27\times10^{-4}$ \\ 
             & MVA & $1.74\times10^{-4}$ & $3.50\times10^{-4}$ & $1.47\times10^{-3}$ \\ \hline
             \multirow{3}{*}{$10\%$}& DIM$(t=t_{\gamma})$ & $3.57\times10^{-4}$ & $1.81\times10^{-4}$ & $1.80\times10^{-4}$ \\ 
              & DIM$(t=2t_{\gamma})$ & $8.35\times10^{-4}$ & $3.14\times10^{-4}$ & $4.02\times10^{-4}$ \\
             & MVA & $4.11\times10^{-4}$ & $7.66\times10^{-4}$ & $5.61\times10^{-4}$ \\\hline
        \end{tabular}
        \caption{Vasicek setting fixed $\theta=3\%,\,r_0=1\%,\,\delta=0\%$. Relative errors for DIM at times $t_{\gamma}$ and $2t_{\gamma}$, and relative error in MVA.}
        \label{tab:eVasicek}
    \end{table}

    Based on these combinations, we present in Tables \ref{tab:eVasicek} and \ref{tab:eHW} the relative errors achieved for DIM at the monitoring times $t_{\gamma}$ and $2t_{\gamma}$, as well as in the calculation of the MVA. Broadly speaking, they are presented in such a way that we can identify the higher volatility scenarios with those presented in the upper triangular part of the tables, and the lower ones with those presented in the bottom triangular part. In general, the conclusions drawn for both configurations on the previous experiment are confirmed. The worst-case scenarios for Vasicek are those with combinations of small mean reversion speed and large model volatility, losing about an order of magnitude compared to the opposite cases. In some cases, the MVA values also lose precision compared to the individual DIM predictions, which is expected due to the accumulation of errors at each monitoring time. 
    
    \begin{table}[ht]
        \centering
        \begin{tabular}{|m{1.5cm} m{3cm}|c c c|}
            \hline
            \multirow{2}{*}{$a$} & \multirow{2}{*}{Target} & \multicolumn{3}{c|}{$\sigma$} \\ \cline{3-5} 
             &  & $0.05\%$ & $0.75\%$ & $1.5\%$ \\ \hline \hline
            \multirow{3}{*}{$1\%$} & DIM$(t=t_{\gamma})$ & $4.10\times10^{-4}$ & $1.41\times10^{-5}$ & $2.65\times10^{-4}$ \\ 
             & DIM$(t=2t_{\gamma})$ & $2.92\times10^{-5}$ & $4.27\times10^{-5}$ & $2.64\times10^{-4}$ \\ 
             & MVA & $2.92\times10^{-4}$ & $1.86\times10^{-4}$ & $7.60\times10^{-5}$ \\ \hline
            \multirow{3}{*}{$2.5\%$} & DIM$(t=t_{\gamma})$ & $1.86\times10^{-4}$ & $2.22\times10^{-4}$ & $1.78\times10^{-4}$ \\ 
             & DIM$(t=2t_{\gamma})$ & $2.73\times10^{-5}$ & $1.46\times10^{-4}$ & $5.94\times10^{-5}$ \\ 
             & MVA & $9.30\times10^{-5}$ & $1.81\times10^{-4}$ & $2.27\times10^{-4}$ \\ \hline
             \multirow{3}{*}{$5\%$}& DIM$(t=t_{\gamma})$ & $2.21\times10^{-4}$ & $2.46\times10^{-4}$ & $1.81\times10^{-4}$ \\ 
              & DIM$(t=2t_{\gamma})$ & $5.91\times10^{-5}$ & $5.92\times10^{-5}$ & $2.45\times10^{-4}$ \\
             & MVA & $1.68\times10^{-4}$ & $3.08\times10^{-4}$ & $2.06\times10^{-5}$ \\\hline
        \end{tabular}
        \caption{Hull-White setting fixed $\beta_0=1\%,\,\beta_1=\beta_2=0.5\%,\,\delta=0\%$. Relative errors for DIM at times $t_{\gamma}$ and $2t_{\gamma}$, and relative error in MVA.}
        \label{tab:eHW}
    \end{table}

    In the case of Hull-White, however, the situation is different. The error distribution does not show the pattern observed in Vasicek. It exhibits a more random behavior across the combinations of $a$ and $\sigma$. This is explained by the fact that these errors are very close to, or even below, the MC variance threshold used for simulating the ground truth (reference) values. All in all, the obtained estimations by the DIM learning approach are very satisfactory since, for both models, they (almost) achieve the MC precision level, at much less computational cost.

    \subsection{Portfolio application}
    This last subsection is intended to demonstrate how well the methodology generalizes when applied to a practical example of major relevance. For this purpose, we define a portfolio consisting of $6$ ATM interest rate swaps with maturities $T_{\phi} = 5 +\phi,\,\,\phi=0,\dots, 5$. Those defined by a even index are payer swaps, while the rest are receiver swaps. In addition, the first three have a quarterly vs. semi-annual payment schedule, and the others a semi-annual vs. annual schedule. All of them are defined on a notional of $100$ monetary units.

    \begin{figure}[ht]
        \centering
        \begin{subfigure}[b]{0.49\textwidth}
            \centering
            \includegraphics[page=1, width=\textwidth]{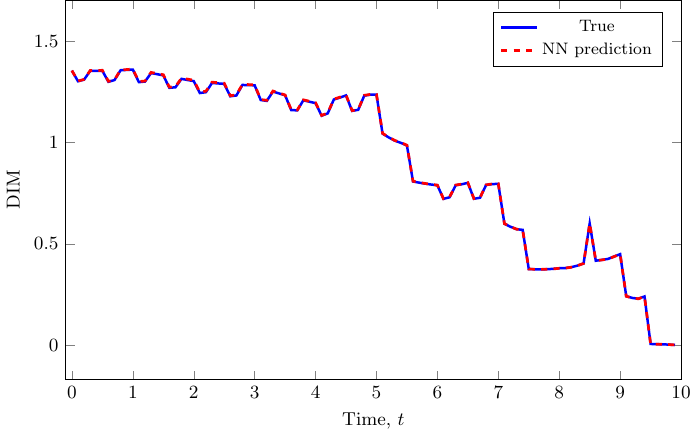}
        \end{subfigure}
        \hfill
        \begin{subfigure}[b]{0.49\textwidth}
            \centering
            \includegraphics[page=2, width=\textwidth]{Plots/swapPortfolioTrueVsPred.pdf}
        \end{subfigure}
    \caption{Samples of DIM portfolio prediction (dashed red line) versus the ground truth trajectory (blue line). \underline{Left}: Vasicek setting for the initial market state $a=7.23\%$, $\sigma=1.1\%$, $\theta=1.2\%$, $r_0=1.4\%$. \underline{Right}: Hull-White setting for the initial market state $a=4.13\%$, $\sigma=1.2\%$, $\beta_0=4.93\%$, $\beta_1=0.29\%$, $\beta_2=0.91\%$.}
    \label{fig:predVsTrueDIM}
    \end{figure}

    Most of the dataset simulation details are kept. The training set consists of $K_T=2^{22}$ IM sampled labels from the above-mentioned market state settings, see Table \ref{tab:ParamsSetting}. The only difference is that we omit the swap spread parameter, since now it has nonsense. In addition, we consider $N=100$ equally spaced monitoring times in which DIM values is intended to be known. These changes lead to a neural network architecture with one less input parameter in each setting, and which $100$ outputs. No additional changes are made to the other features of the network. Lastly, a test set with $K_V=32$ samples is generated with the same number of MC paths as before.

    In order to evaluate the stability and the performance of the methodology, we run $10$ training trials for each setting, preserving the optimiser and the training strategies used before. In relation to the global validation metrics, we get RMSE $95\%$ confidence intervals of $3.439\times10^{-3}\pm 7.19\times10^{-4}$ and $2.373\times10^{-3}\pm 9.71\times10^{-4}$ for Vasicek and Hull-White, respectively. These are positive results, in line with those shown for the simplified case, and demonstrate stability in training convergence.

    For completeness, we randomly select an initial market state from each simulated ones (and for each setting) and study the network performance in MVA computation for such a particular case. The $95\%$ confidence interval for the MVA relative errors are $1.912\times10^{-3}\pm 1.11\times10^{-3}$ and $1.845\times10^{-3}\pm7.28\times10^{-4}$ for Vasicek and Hull-White, respectively. The results seem to confirm the same behavior as in the previous case study, the higher variance of the Vasicek model is manifested in higher errors. However, making the portfolio more complex with a larger number of assets does not hamper the performance, which is very positive. This situation is reflected in all the others samples and trial executions, since the maximum MVA relative error is $6.753\times10^{-3}$ for Vasicek and $3.96\times10^{-3}$ for Hull-White. Lastly, we show in Figure \ref{fig:predVsTrueDIM} the approximation of the DIM trajectory (compared to the ground truth value), for each setting, estimated by means of one of the trained networks. The initial market states are also specified. As we can see, the approximation is qualitatively indistinguishable from the reference, expected from what we have described.

 \label{sec:Results}
\section{Conclusion} \label{sec:Conclusion}

One of the main disadvantages of training neural networks by means of classical supervised training is that the elaboration of the training dataset is highly expensive, specially in the kind of task where nested MC simulation are required. The main idea of this work has been to address this problem for the DIM computation, a CCR quantity of interest due to its role in the MVA. By definition, DIM is calculated from the expectation of IM over the life of the portfolio to which it is related, and the nested behaviour appears if one wants to compute it. In this sense, training neural networks for DIM approximation in a classical supervised fashion make their implementation in an online CCR engine unfeasible. Since the DIM framework fits perfectly with the sample payoffs training presented by the authors of \cite{Savine2020}, in this work we extend their idea to the neural network computation of the quantity in question. 

First, we have condensed all the information required for its calculation into a vector of initial market state variables at initial time. This has been achieved by means of an interest rate evolution model, unequivocally determined from a few parameters, together with a parsimonious parameterisation of the current interest rate term structure. From the initial market state vector, which is the set of input values for our neural network model, we build a training dataset in which the labels are not ground truth DIM values, but highly noisy and unbiased samples of DIM generated with a single MC path, i.e., we use IM realizations as DIM approximations. Note that the DIM for a given initial state is a function of time. We have handled this particularity by defining a mesh of monitoring times in which training and validation values are defined, which means that our neural network model has a multi-output structure. This choice allows for a simple extension of the concept of conditional expectation, as well as greatly facilitating the training process. The proposed approach presents two important advantages. First, for the price of a single MC execution we can build an entire dataset ready for the network training, and second, our trained network has been parameterised by the initial market state variables, so, once trained, we have a neural network model that approximates DIM trajectories at monitoring times for a whole set of initial states, largely avoiding re-training.

Having described the introduced methodology, we have presented convergence and sensitivity results of the approximations based on the initial data for a simple example, a portfolio consisting of a single swap, and for two interest rate models, Vasicek and Hull White. From these experiments, we can extract a couple of main conclusions. Firstly, there is a convergence ratio depending on the number of training samples used. This is worse than the square root order reduction given by MC, in exchange for having parameterised a model as a function of the initial conditions. Secondly, the obtained DIM estimations, and they are not particularly affected by the input data near initial state boundary domain. Finally, we have found that the methodology performs well in a portfolio of swaps, a more realistic scenario, highlighting the general applicability of the proposed solution.


\section*{Acknowledgements}
The authors' research has been funded by the Spanish MINECO under research project number PDI2019-108584RB-I00. The authors also acknowledge the support of the CITIC research centre. CITIC is funded by the Xunta de Galicia through the collaboration agreement between the Consellería de Cultura, Educación, Formación Profesional e Universidades and the Galician universities for the reinforcement of the research centres of the Galician University System (CIGUS). Joel P. Villarino acknowledge the support received by the Xunta de Galicia under grant ED481A/2023-202.

\bibliographystyle{abbrv}
\bibliography{refs}

\end{document}